\documentclass[aps,pra,twocolumn,groupedaddress]{revtex4-1}
\usepackage{graphicx}
\usepackage{amsmath}
\usepackage{amsfonts}
\usepackage{float}
\usepackage{mathtools}
\usepackage{epstopdf}

\begin{document}


\title{Generation, manipulation, and detection of two-qubit entanglement in waveguide QED}


\author{C. Gonzalez-Ballestero}
\author{Esteban Moreno}
\email{esteban.moreno@uam.es}
\author{F.J. Garcia-Vidal}
\affiliation{Departamento de F\'isica Te\'orica de la Materia Condensada, and Condensed Matter Physics Institute (IFIMAC), Universidad Aut\'onoma de Madrid, Madrid 28049, Spain}


\date{\today}

\begin{abstract}
We study the possibility of using guided photons to generate, control, and measure the entanglement of two qubits that is mediated by a one-dimensional waveguide. We show how entanglement can be generated both with single photon and with two-photon wavepackets. The introduction of a second photon allows for a manipulation of the entanglement between the qubits, and phenomena such as sudden death and revival of entanglement appear.  Finally, we propose a procedure for entanglement detection via the scattering output of a single-photon over a qubit state.
\end{abstract}

\pacs{}

\maketitle

\section{Introduction}

During the last years, a huge research effort has been directed towards the efficient control of the interaction between quantum emitters and the electromagnetic (EM) field. This ability has turned out to be particularly interesting for quantum computation and quantum information, where the development of efficient quantum devices requires a careful control of light-matter interaction at the quantum level. Since quantum entanglement is a fundamental resource for quantum information, an adequate way of controlling, manipulating, and detecting it is extensively sought. Many promising applications of entanglement between light and matter have been proposed, such as single-photon transistors \cite{ChangNATPHYS2007}  or simple quantum logical gates \cite{PlantenbergNATURE2007}. The fundamental concept all these applications rely on is the two-qubit system and the possibility of modifying its properties by coupling the qubits to a structured EM environment. Two main ways have been proposed to achieve this environment modification: optical cavities \cite{WaltherREPPROGPHYS2006}, in which the dynamics is driven by either a single mode or a collection of discrete modes, and waveguides \cite{YaoOPTEXPRESS2009,BermelPRA2006}, where the qubits are coupled to a continuum of modes. Systems based on cavity quantum electrodynamics (QED) have been studied in more detail, as their higher confinement allows for the observation of properties associated to strong light-matter coupling, such as Rabi oscillations \cite{KaluznyPRL1983,BrunePRL1996}. On the other hand, waveguides offer an easy way for the in- and out- coupling of information to the system, which makes them good candidates for a large-scale implementation. Moreover, a large variety of waveguide systems with different properties have been analyzed, including those based on photonic crystals \cite{WongPHOTNANO2009,WasleyAPL2012}, plasmons \cite{ChangPRL1997,DiegoPRB2011,GlezTudelaPRL2011,AkimovNATURE2007,DzsotjanPRB2010,AndersenNATPHYS2010}, superconductors \cite{vanLooSCIENCE2013,HoiPRL2011}, dielectric materials \cite{QuanPRA2009,BogaertsLIGHTWAVETECH2005}, and other systems \cite{BensonNATURE2011,MoeferdtOPTLETT2013}. For these reasons, waveguide QED provides an excellent workspace for quantum information applications. 

Previous investigations in waveguide QED have reported the possibility of entanglement generation between qubits coupled to plasmonic waveguides \cite{DiegoPRB2011,GlezTudelaPRL2011,DzsotjanPRB2010}, and its applicability for the design of simple quantum logic gates. The analysis in all these works made use of a master equation formalism, in which the EM degrees of freedom are traced out. In order to keep track of the guided modes' evolution, a real-space formalism has been developed \cite{FanOptLett2001} and successfully used in single-photon \cite{FanPRA2009} and two-photon scattering problems \cite{BarangerPRA2010,BarangerPRL2013,RoyPRB2010}. Recently, we have applied this approach to the problem of entanglement generation \cite{GonzalezBallesteroNJPHYS2013}. Our proposal required an out-of-equilibrium initial state, in which the qubits were prepared in a particular configuration. The preparation of such initial state is not always an easy task in the experiment, and this motivates the search for alternative entangling schemes.

In the present work we explore new ways to generate, manipulate, and detect the entanglement by exploiting the simple input and output scheme offered by waveguides. In the first part of the paper we study the ability of guided photons to generate entanglement between two qubits separated by a short distance. We show that, although one single photon can accomplish this task, the addition of a second photon introduces new interesting phenomena such as sudden death and revival of the entanglement. These features can be exploited to manipulate the entanglement dynamics of the two qubits in a simple way by varying the temporal delay between the two incident photons. In the second part of the work, we address the problem of detecting the entanglement between the qubits in waveguide QED systems. We show that, when a single photon probe impinges on the two-qubit system, the scattered photonic state contains information about the two-qubit initial entanglement.

This paper is organized as follows. In section II we introduce the system, the Hamiltonian, and the single-photon dynamics. In section III we study the effect of including a second photon into the picture. Next, we address the problem of entanglement detection in section IV. The conclusions are presented in section V. Finally, the Appendix is devoted to the detailed diagonalization of the Hamiltonian in the two-excitation subspace.

\section{Single-photon dynamics}

\begin{figure}
  \begin{center}
    \centerline{\includegraphics[scale=0.23]{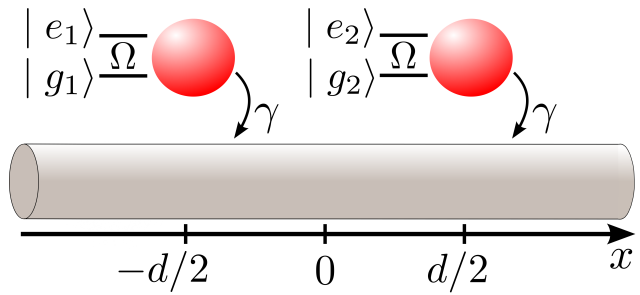}} 
    \vspace{0cm}
  \caption{(Color online) Illustration of the system under study. Two qubits with ground state $\vert g \rangle$ and excited state $\vert e \rangle$ are placed in the vicinity of a waveguide. Both have a transition frequency $\Omega$ and an individual decay rate $\gamma$.\label{fig:system}}
  \end{center}
  \vspace{-0.5cm}
\end{figure}

The system under study consists of two qubits side-coupled to a waveguide, as depicted in Fig.~\ref{fig:system}. The EM field in the waveguide is modeled as a one-dimensional, continuous collection of bosonic modes with linear dispersion relation $\omega = v_g \vert k\vert $, $\omega$ being the angular frequency, $k$ the wavevector, and $v_g$ the group velocity, which we will choose equal to $c$ for simplicity. The waveguide is set along the $x$ axis and the propagation losses are assumed to be negligible. The qubits, labeled as $1$ and $2$, are two-level systems with transition frequency $\Omega$, interacting with the guided modes with a coupling energy $\gamma$. They are placed close enough to each other such that we can assume the qubits-photon coupling to occur at the same position $x=0$, i.e., the distance between the qubits being $d << \lambda = 2\pi v_g/\Omega$. This approximation simplifies the problem as the qubits will now be coupled only to photonic states with even parity. The separation between the qubits, however, is chosen to be large enough to suppress the direct coupling between them. In this way, the qubit-qubit interaction will be completely mediated by the guided modes. Our aim is to model systems where the decay channel for the qubits is basically the coupling to guided photons. We will therefore neglect any additional radiative losses of the qubits.

As mentioned in the introduction, previous works \cite{DiegoPRB2011,GlezTudelaPRL2011} have used a Markovian master equation to study the dynamics of the two-qubit system. These studies show  that the waveguide-mediated interaction induces the formation of two collective entangled states, the sub- and superradiant states. Such behavior has been recently observed in experiments \cite{vanLooSCIENCE2013}. In this work, we are interested in entanglement generation, manipulation, and detection schemes based on the interaction between the qubits and guided modes whose properties can be modified in a controlled way. The formalism required for this purpose must include the photonic degrees of freedom in detail. Hence, we will make use of the following real-space Hamiltonian \cite{FanOptLett2001,FanPRA2009}:
\begin{equation} \label{HRL}
\begin{aligned}
H =-iv_g\int dx \big(c_R^\dagger(x)\partial_x & c_R(x) - c_L^\dagger(x)\partial_xc_L(x)\big) + \\ + \Omega\sum_{i=1,2} & \sigma_i^\dagger\sigma_i + \\
 +\frac{V}{2} \sum_{i=1,2}\int dx \delta(x) \left[\left(c_R^\dagger\!\!\right.\right. & \left.\left.\!\!(x) +c_L^\dagger(x)\right)\sigma_i + \text{h.c.}\right].
\end{aligned}
\end{equation}
($\hbar = 1$). The first term is the Hamiltonian of the waveguide modes, which are characterized by creation operators $c_{R/L}^\dagger(x)$. These bosonic operators create a right/left propagating photon at the position $x$. The second term contains the energy of the qubits expressed in terms of the fermionic creation and annihilation operators $\sigma_i^\dagger = (\sigma_i)^\dagger=\vert e_i\rangle\langle g_i\vert$, the states $\vert g_i\rangle$ and $\vert e_i\rangle$ being the ground and excited states of the qubit $i$, respectively (see Fig. \ref{fig:system}). The third term describes the qubit-photon interaction that takes place at $x=0$ and is characterized by a coupling strength $V$. The coupling constant in units of energy is given by $\gamma = V^2/v_g$.

Following the standard procedure, we take advantage of the inversion symmetry by defining the even and odd operators as $c_{e,o}(x) = \left(c_R(x)\pm c_L(-x)\right)/\sqrt{2}$ ; $\sigma_{e,o} = \left(\sigma_1\pm \sigma_2\right)/\sqrt{2}$. In this basis the Hamiltonian reads
\begin{equation}\label{HEO}
\begin{aligned}
H& = -iv_g\int dx \big(c_e^\dagger(x)\partial_xc_e(x) + c_o^\dagger(x)\partial_xc_o(x)\big) +\\+& \Omega\left( \sigma_e^\dagger\sigma_e + \sigma_o^\dagger\sigma_o\right) + V\int dx \delta(x) \left[c_e^\dagger(x)\sigma_e + \text{h.c.}\right].
\end{aligned}
\end{equation}
In Eq. (\ref{HEO}), no odd symmetry operator appears in the interaction terms as we work in the limit $d << \lambda = 2\pi v_g/\Omega$. The odd symmetry qubit state, $\sigma_o^\dagger\vert 0\rangle$, is then uncoupled to the waveguide modes. Thus, it can be identified with the subradiant state mentioned above. Physically, this implies that for small qubit-qubit separations the even modes play the main role in the system dynamics, and only the superradiant state $\sigma_e^\dagger\vert 0\rangle$ interacts with the photonic modes. 

The diagonalization of the Hamiltonian in the one-excitation subspace has been carried out in Refs. \cite{FanPRA2009,GonzalezBallesteroNJPHYS2013}, where the general time-evolution operator has been obtained. By using this result we can calculate any time-dependent expected value for a single-excitation initial state. We will choose the single-photon initial state
\begin{equation}\label{singlePSI0}
\vert \Psi (0) \rangle = \int dx \; \psi (x;x_0) c^\dagger_R(x) \vert 0 \rangle.
\end{equation}
The shape of this wavepacket is defined as
\begin{equation}\label{Psia}
\psi(x;a) \equiv \sqrt{2\mu/v_g}\cdot e^{\mu a/v_g}e^{-\mu \vert x \vert/v_g}e^{i\omega x/v_g}\theta(-x-a).
\end{equation}
This particular shape is sensible from the theoretical point of view, as the sharp wavefront makes the qubit-photon interaction to start at a well defined time. Hence, the interpretation of the results is more transparent as we know exactly when the photons arrive at the position of the qubits. Note that this wave profile is the same as the one generated by the decay of an additional qubit with decay rate $\mu$ and frequency $\omega$, behaving as a photon source. If we assume this is the case, one has to take into account that the reflected pulse would interact with the source qubit, requiring an extension of the formalism. Such interaction could be avoided by insertion of a circulator, at the cost of increased noise in the system. Nevertheless, if the loss in the waveguide is low, as we assumed from the beginning, the source can be located very far from the qubit pair and this effect should not play a very significant role.


In the following calculations, the frequency $\omega$ in the incident wavepacket [Eq. (\ref{Psia})] is set to $\omega = \Omega$. It can be checked that this value maximizes the photon-qubit interaction, as the incoming probability distribution shows a maximum in $\omega$ in the frequency space. The qubits-waveguide coupling has been chosen as $\gamma = \Omega/100$, which is within the experimental range for superconducting systems \cite{NiemczykNATPHYS2010}. Nevertheless, we have checked that the features of the generated entanglement are not very sensitive to the particular value of $\gamma$. Instead, the time-evolution of the entanglement depends on the ratio $\gamma/2\mu$, as we will show below. Finally, we place the $t=0$ wavefront at $x_0 = 200\lambda$. This choice is arbitrary and does not change the system dynamics, as the waveguide is not dispersive.

Since we are interested in the amount of entanglement generated in the two-qubit system, we must introduce a magnitude to quantify the entanglement between both qubits. Among several proposals, Wootters Concurrence \cite{WoottersPRL1998} has turned out to be the most appropriate choice for two-qubit systems. This quantity is a continuous function that ranges from $0$, for completely disentangled states such as $\vert g_1g_2\rangle$ or  $\vert e_1e_2\rangle$, to $1$ for maximally entangled states like $\vert +\rangle = \sigma_e^\dagger\vert 0\rangle$ and $\vert -\rangle = \sigma_o^\dagger\vert 0\rangle$. The calculation of the concurrence, $C$, requires the reduced density matrix of the two-qubit system, which is calculated by taking the partial trace of the total density matrix \cite{Louisell}. For the initial state in Eq. (\ref{singlePSI0}), the concurrence turns out to be equal to the population of the even state $\vert + \rangle$, i.e., $C = \rho_+(t)$.

\begin{figure}
  \begin{center}
    \centerline{\includegraphics[scale=0.32]{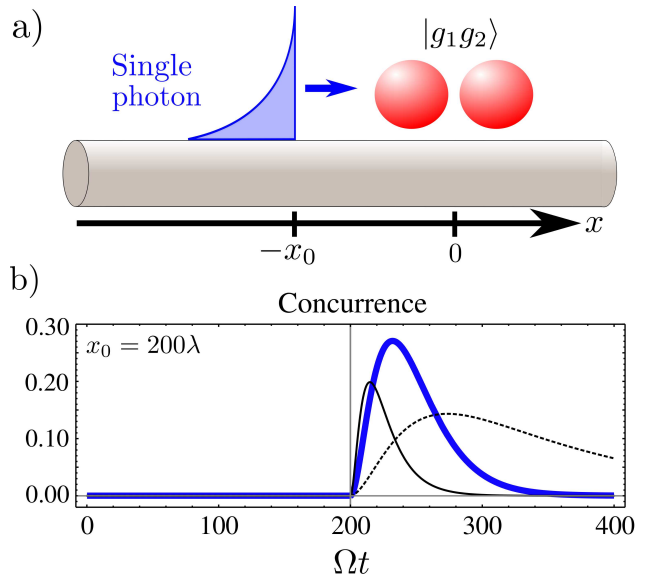}} 
    \vspace{0cm}
  \caption{(Color online) Entanglement generation via a single-photon scattering. a) Scheme of the initial state: a single photon at a distance  $x = -x_0$ and with the profile of Eq. (\ref{Psia}) impinges on the two qubits in their ground state. b) Concurrence of the two-qubit system as a function of time, for $\mu = \gamma/2$ (blue thick line), $\mu = 2\gamma$ (thin solid line), and $\mu = \gamma/15$ (dashed line). \label{fig:2}}
  \end{center}
    \vspace{-1.0cm}
\end{figure}

By applying the time-evolution operator to the initial state in Eq. (\ref{singlePSI0}), we can analytically obtain the concurrence of the two-qubit system. The result is shown in Fig. \ref{fig:2}(b). We observe how the qubit-photon interaction is turned on when the photon arrives at the qubits position, $x = 0$. The highest value for the concurrence is achieved for $\mu = \gamma/2$ (blue line), and the curve rises up to $C \approx 0.27$. This is quite a high value considering the low symmetry of the initial state. In the same panel, the concurrence for the cases $\mu = 2\gamma$ (solid black line) and $\mu = \gamma/15$ (dashed black line) is shown. The entanglement generation is clearly worsened for both situations, as a result of the inefficient excitation of the qubits. Indeed, the spectral linewidth of the pulse in Eq. (\ref{Psia}), namely $2\mu$, represents the rate at which the single excitation is arriving at the qubits, whereas $\gamma$ quantifies the photon-qubits interaction rate, i.e., the rate at which qubits can absorb the photon. Hence, the interaction is maximized when these two rates are equal, $\gamma/2\mu =1$, as mentioned above.

\section{Two-photon dynamics: manipulating the entanglement}

In this section we address the problem of manipulating the entanglement, i.e., modifying the shape of the curve in Fig. \ref{fig:2}(b). This could be achieved by varying the parameters $\lbrace\mu,\omega\rbrace$ of the incident wavepacket [Eq. (\ref{Psia})]. This method, however, is experimentally challenging as it demands a single-photon tunable source. Here, we propose to introduce a wavepacket formed by two identical photons instead of a single one. This approach does not require a modification of the photon source, as it is a natural extension of the single-photon problem.

\begin{figure}
  \begin{center}
    \centerline{\includegraphics[scale=0.3]{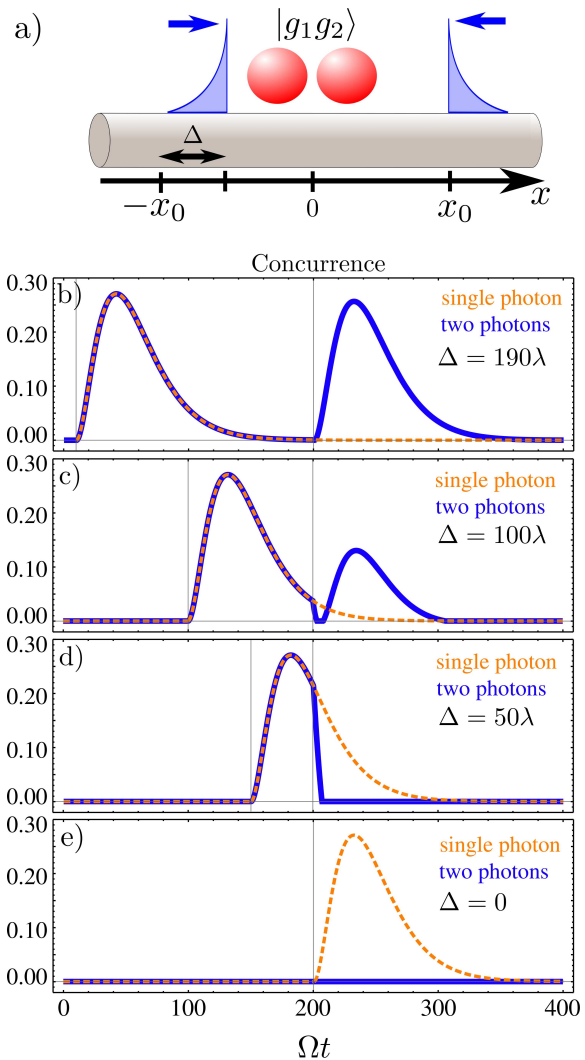}} 
    \vspace{0cm}
  \caption{(Color online) Two-photon situation. a) Scheme of the initial state: two single photons impinge over the qubits in their ground state, one from the left and the second from the right. The initial positions of the photons are $-(x_0-\Delta)$ and $x_0$, respectively. b)-d) Time evolution of the concurrence, compared with the single-photon generation studied in previous section, for different photon-photon delays $\Delta$. \label{fig:3}}
  \end{center}
  \vspace{-1.0cm}
\end{figure}

In order to work with a two-photon initial state, the diagonalization of the Hamiltonian of Eq. (\ref{HEO}) has been carried out in the two-excitation Fock space (see Appendix). The initial state is schematically depicted in Fig. \ref{fig:3}(a): it is formed by one photon coming from $x=-\infty$ and propagating rightwards, arriving to the two qubits earlier than a second photon originating from $x=+\infty$ and propagating leftwards. By choosing the propagation quantum numbers $(R,L)$ to be different we make sure that the photons are distinguishable and thus the symmetrization of the initial state is not necessary. The complete expression of this two-photon state is given by
\begin{multline}\label{GeneralPSI0simpler}
\vert \Psi (0) \rangle = \int dx_1 \; \psi (x_1;x_0-\Delta) c^\dagger_R(x_1) \cdot \\ \cdot\int dx_2  \; \psi (-x_2;-x_0) c^\dagger_L(x_2)\vert 0 \rangle,
\end{multline}
where the wavepacket $\psi(x;a)$ has been defined in Eq.~(\ref{Psia}). The first photon arrives at the position of the qubits at a time $\Delta/v_g$ earlier than the second. We will keep the same values for the parameters $\lbrace\gamma,\omega\rbrace$ as in the previous section ($\gamma=\Omega/100;\;\omega=\Omega$), and set $\mu = \gamma/2$ in order to maximize the generated entanglement. The arrival time of the second photon is also fixed through its initial position, $x_0 = 200\lambda$. We will then explore the dependence of the system evolution with the delay $\Delta$, which determines the arrival time of the first photon.

In this case, the partial trace over the total density matrix has also been performed, and the calculation of the concurrence yields the expression:
\begin{equation}\label{concurrence}
C(t) = \text{max} \left[ 0, \rho_+-2\sqrt{\rho_\beta\rho_{GS}}\right],
\end{equation}
where the quantities $\lbrace \rho_+, \rho_\beta, \rho_{GS}\rbrace$ are the populations of the qubit states $\lbrace \vert + \rangle, \vert e_1 e_2\rangle, \vert g_1 g_2\rangle \rbrace$, respectively. These three states, together with the uncoupled state $\vert - \rangle$, form a basis of the two-qubit system. The amount of entanglement is thus given by two competing terms: it grows as the population of the even, entangled state $\vert + \rangle$ increases, while decreasing when the populations of the disentangled states $\vert e_1 e_2\rangle, \vert g_1 g_2\rangle$ build up.

Each of the populations in Eq. (\ref{concurrence}) is analytically obtained from the time evolution of the initial state, Eq. (\ref{GeneralPSI0simpler}). The concurrence generated by this two-photon wavepacket is depicted in Figs. \ref{fig:3}(b) - \ref{fig:3}(e) (solid line), compared to the single-photon case from previous section (dashed line). For a large delay $\Delta$ [panel \ref{fig:3}(b)], the two-photon curve is very similar to the sum of two similar single-photon pulses. This is due to the fact that the two-qubit system has time to relax to its ground state before the arrival of the second photon. When the delay is reduced, the two pulses get closer and start to interact in a complex way as seen in panels \ref{fig:3}(c) and \ref{fig:3}(d). Here, the arrival of the second photon produces a very rapid decay of the entanglement which, in the case of panel \ref{fig:3}(c), is followed by a rebirth after a given time. Finally, when the two photons arrive to the qubits at the same time [panel \ref{fig:3}(e)] no entanglement is generated. This means that, regarding the concurrence, the second photon  fully suppresses the effect of  the first one.

\begin{figure}
  \begin{center}
    \centerline{\includegraphics[scale=0.27]{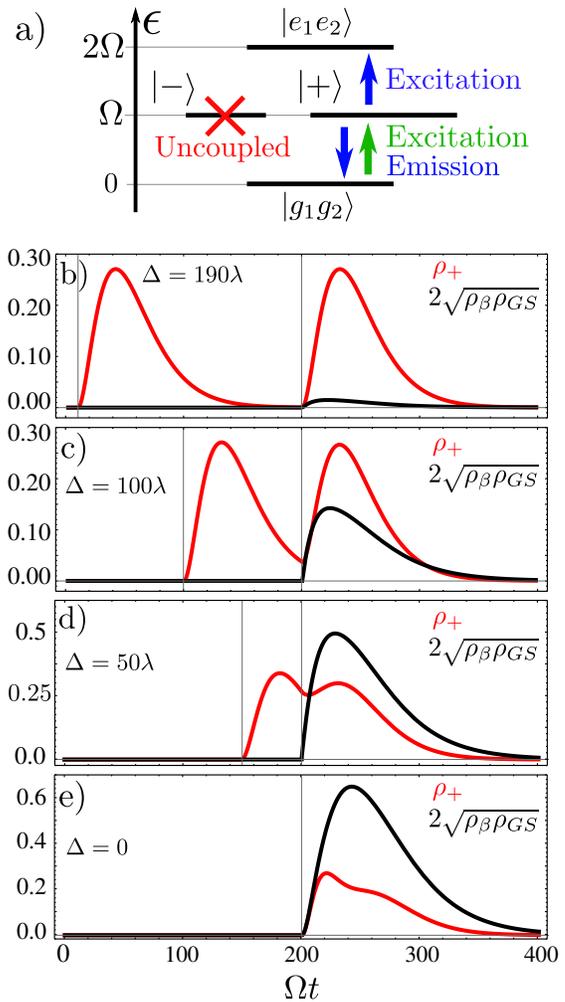}} 
    \vspace{0cm}
  \caption{(Color online) a) Level structure of the two-qubit system. The subradiant state is uncoupled from the waveguide and does not interact with the guided photons. b)-e) Time evolution of the two competing terms in the expression of the concurrence [Eq. (\ref{concurrence})], for different photon-photon delays $\Delta$. \label{fig:4}}
  \end{center}
  \vspace{-1cm}
\end{figure}

The interesting dynamics shown in Fig. \ref{fig:3} can be understood in terms of the population dynamics of the qubit levels, which are depicted schematically in Fig. \ref{fig:4}(a). Here, the three possible photon-induced transitions are also shown: the excitation processes $\vert g_1 g_2\rangle \to \vert +\rangle$ and $\vert +\rangle \to \vert e_1 e_2\rangle$, and the induced emission $\vert + \rangle \to \vert g_1 g_2\rangle$. Note that the spontaneous emission processes, $\vert e_1 e_2\rangle \to \vert +\rangle$ and $\vert +\rangle \to \vert g_1 g_2\rangle$, are not shown in the picture. The competition between the two  terms in Eq. (\ref{concurrence}) accounts for the concurrence evolution. Figures \ref{fig:4}(b) - \ref{fig:4}(e) show both of these terms, $\rho_+(t)$ and $2\sqrt{\rho_\beta(t)\rho_{GS}(t)}$, for the same values of $\Delta$ considered in Fig. \ref{fig:3}. For large delay [panel \ref{fig:3}(b)], the state $\vert + \rangle$ excited by the first photon decays almost completely before the arrival of the second one. The qubits' states probed by the first and second photon are approximately the same, namely $\vert g_1g_2\rangle$. As a consequence, the evolution of the concurrence with time displays two similar peaks. When the delay is shortened as in panels \ref{fig:4}(c) and \ref{fig:4}(d), the population of the state $\vert + \rangle$ is significant at the arrival of the second photon. In this situation, the transitions from this state, marked as blue arrows in Fig. \ref{fig:4}(a), start to be relevant and thus the square root term in Eq. (\ref{concurrence}) increases. Eventually, the condition $2\sqrt{\rho_\beta(t)\rho_{GS}(t)} \ge \rho_+(t)$ is satisfied. At this point the concurrence vanishes in a phenomenon known as \textit{sudden death of entanglement} \cite{Tanas2}. If the population of the state $\vert + \rangle$ at the arrival of the second photon is not too high [panel \ref{fig:4}(c)], the square root term only becomes slightly larger than $\rho_+(t)$. As a consequence, the transitions $\vert e_1 e_2 \rangle \to \vert + \rangle$ and $\vert g_1 g_2 \rangle \to \vert + \rangle$ sufficiently increase the population of the $\vert + \rangle$ state so as to produce a \textit{revival of entanglement}. Such phenomenon, however, does not take place in panel \ref{fig:4}(d) where the two curves are far apart from each other. In the zero-delay limit [panel \ref{fig:4}(e)], the concurrence does never rise, as the condition $2\sqrt{\rho_\beta(t)\rho_{GS}(t)} \ge \rho_+(t)$ is fulfilled during all the time evolution. This is a consequence of the maximally efficient pumping of the doubly excited state $\vert e_1 e_2 \rangle$. Entanglement sudden death and revival are well-known processes in quantum optics, which have been theoretically studied \cite{YuPRL2004,ZyczowskiPRA2001} and experimentally observed \cite{XuPRL2010,AlmeidaScience2007} in cavity QED and other systems, although, up to our knowledge, not in waveguides.

 The analysis of Figs. \ref{fig:3} and \ref{fig:4} shows that we can easily tune the entanglement generation by means of the second photon delay: when selecting large values of $\Delta$, photons act independently so we can generate a train of pulses in the concurrence. However, by decreasing the delay time we are able to shorten the pulse to the desired duration via sudden death of entanglement. Finally, for zero delay we are able to rise the population of the qubit levels without generating any entanglement. This manipulation of the single-photon concurrence could turn out to be a powerful tool for quantum computation purposes.

\section{Entanglement detection}

So far we have studied the feasibility of generating entanglement between the qubits, as well as manipulating its time evolution. The relevance of these results relies on the possibility of the experimental detection of entanglement. Different schemes have been proposed to measure entanglement both in superconducting qubits \cite{ChowPRA2010} or ion trap systems \cite{RomeroPRA2007}, but they usually require complex quantum operations being performed over the state of the qubits. Moreover, the study of an efficient way of detecting entanglement in waveguide QED systems is lacking. Here, we address this problem by analyzing the scattering of a single photon over a pure qubit state. With this method, we provide a detection procedure that is especially suitable for waveguides, and experimentally affordable as the state of the qubits does not need to be modified. 

\begin{figure*}
  \begin{center}
    \centerline{\includegraphics[scale=0.38]{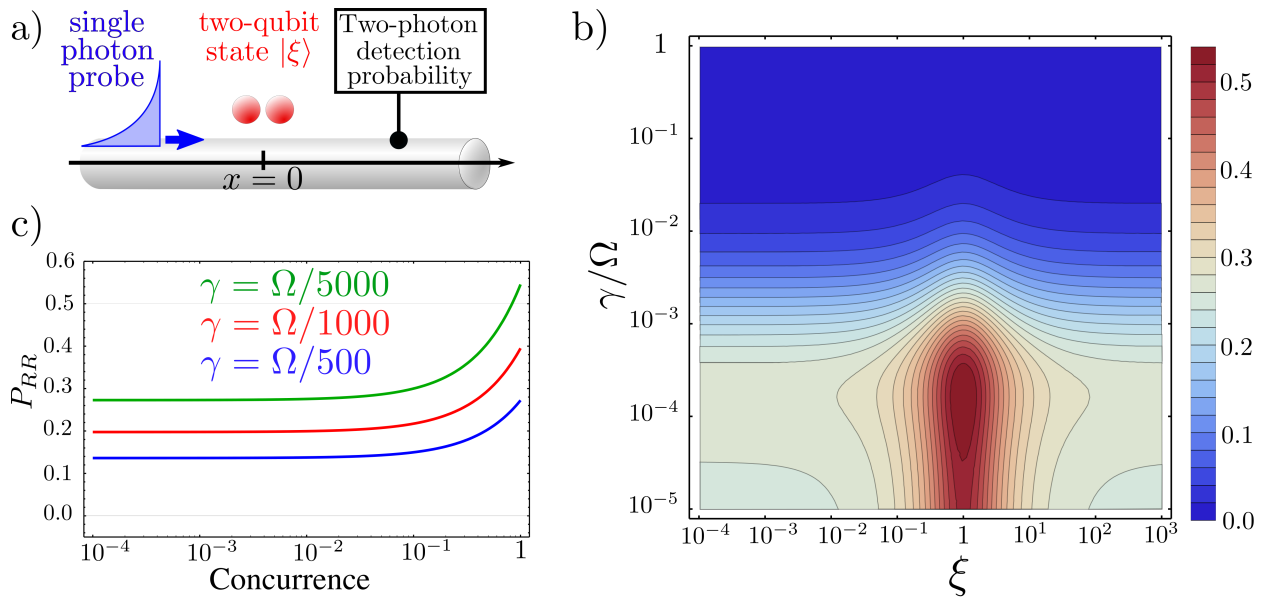}} 
    \vspace{0cm}
  \caption{(Color online) Entanglement detection. a) Scheme of the initial state: a single photon with the shape in Eq.(\ref{Psia}) impinges over the qubits, which are in the single-excitation state parametrized by $\xi$ [Eq. (\ref{Initialdetection})]. A detector is placed at $x > 0$. b) Two-photon total detection probability, $P_{RR}$, as a function of the qubit-waveguide coupling and the entanglement parameter $\xi$. c) Two-photon detection probability as a function of the initial concurrence, for different values of the qubit-waveguide coupling $\gamma$.  \label{fig:5}}
  \end{center}
  \vspace{-0.6cm}
\end{figure*}

The configuration we propose is characterized by an initial state as depicted in Fig. \ref{fig:5}(a): a single-photon with the shape of Eq. (\ref{Psia}) impinges from $x = -\infty$ over the two qubits, which are in a pure, single-excitation state. The expression corresponding to this initial state is
\begin{equation}\label{Initialdetection}
\vert \Psi(0)\rangle = \left(\int dx \; \psi(x;x_0)c_R^\dagger(x)\right)\!\!\otimes\!\! \left(\frac{\sigma_1^\dagger + \xi \sigma_2^\dagger }{\sqrt{1+\vert\xi\vert^2}}\right)\!\vert 0\rangle.
\end{equation}
The parameter $\xi$ ranges from $-\infty$ to $\infty$ and fully determines the state of the qubits. In particular, the disentangled states $\vert e_1 g_2\rangle$ and $\vert g_1 e_2\rangle$ are obtained for $\xi = 0$ and $\xi = \infty$, respectively. Similarly, the even and odd states $\vert \pm \rangle$ correspond to $\xi = \pm 1$. The parameter $\xi$ also determines the entanglement properties of the system. Indeed, the initial concurrence associated with this single-excitation two-qubit state is given by
\begin{equation}\label{ConcXI}
C(\xi) = \frac{2\vert\xi\vert}{1+\vert \xi\vert^2}.
\end{equation}

In order to determine the initial concurrence $C(\xi)$ of the qubits from scattering data we set a detector at some position $x > 0$, as seen in Fig. \ref{fig:5}(a). This detector measures the total probability of collecting two photons propagating rightwards, $P_{RR}$. Our goal is to relate this scattering output to the initial concurrence. We keep the detector open during all the time evolution. This means that the total probability $P_{RR}$ can be obtained by integrating the corresponding probability density over all positive positions:
\begin{equation}\label{PRR1}
\begin{aligned}
P_{RR} & = \\ =\lim_{t\to\infty} &\int_0^\infty \!\!\! dx_1\int_{x_1}^\infty \!\!\! dx_2 \langle c_R^\dagger(x_2) c_R^\dagger(x_1)c_R(x_1)c_R(x_2)\rangle.
\end{aligned}
\end{equation}
The limit $t\to\infty$ is taken in order to guarantee that the photon-qubit interaction has finished, and thus all probability has been taken into account. The lower limit in the first integral, $x_1$, avoids double counting of states. To ensure that the qubit state has not decayed before it has been probed, we have assumed the initial condition $\gamma x_0 << v_g$, i.e., we consider the initial single-photon wavefront to be very close to $x = 0$. 

The simple structure of the initial wavepacket in Eq.~(\ref{Psia}) allows for an analytical calculation of the probability $P_{RR}$, which is displayed in Fig. \ref{fig:5}(b). Here we have set $\omega = \Omega$ as in previous sections, and plotted the two-photon detection probability $P_{RR}$ versus the qubit-waveguide coupling and the entanglement parameter $\xi$. As $\gamma$ is not fixed, the particular value of $\mu$ is now relevant, and we choose it to be smaller than in previous sections, $\mu = \Omega/3000$. In this way we work within the experimental range for many waveguide QED systems \cite{LounisREPPROGPHYS2005}. In Fig. \ref{fig:5}(b) we observe that $P_{RR}$ depends on the parameter $\xi$ and, therefore, on the initial entanglement of the qubits. Note that this dependence is more pronounced for $\gamma = \mu/2$, where $P_{RR}$ reaches a maximum as the decay/excitation rate of the two-qubit system is equal to the rate at which the pulse is arriving. Fig. \ref{fig:5}(c) depicts the two-photon detection probability as a function of the initial concurrence of the qubits, for several values of the qubit-waveguide coupling $\gamma$. These curves show that, for a given $\gamma$, there is a biunivocal relation between $P_{RR}$ and the initial concurrence. 

Although the detection scheme provided above is very clear, its dependence on all system parameters makes it difficult to implement experimentally. In order to use this method we need to know the qubit-waveguide coupling $\gamma$ and transition frequency $\Omega$. Moreover, we have to guarantee that the incident single-photon has the shape of Eq.~(\ref{Psia}), with known parameters $\mu$ and $\omega$. Finally, we need a dispersionless waveguide in order not to modify the wavepacket during its propagation. All these conditions show that a parameter-independent method would be much more useful. For this reason, in what follows we derive a very general result that does not depend on all those details. 

\begin{figure*}
  \begin{center}
    \centerline{\includegraphics[scale=0.35]{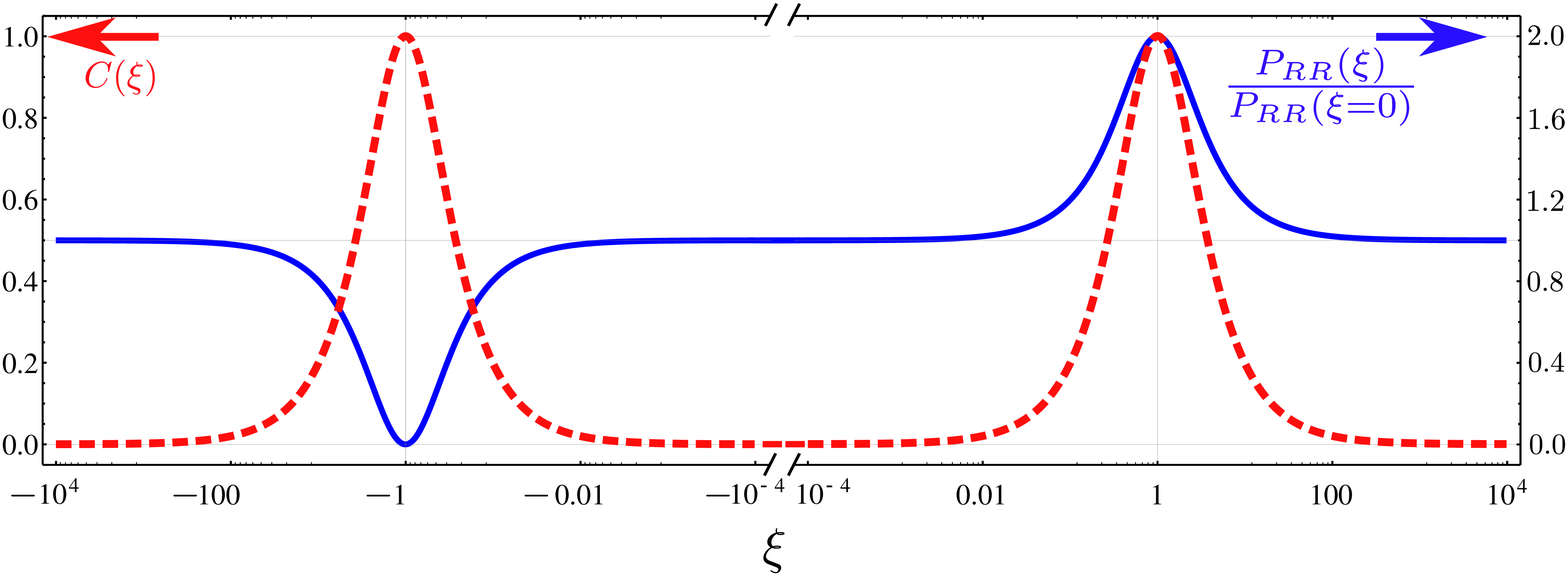}} 
    \vspace{0cm}
  \caption{(Color online) General result for entanglement detection in waveguide QED. The relation between the concurrence of the two qubits (red dashed line) and the normalized two-photon probability (blue solid line) is shown in this picture. \label{fig:6}}
  \end{center}
  \vspace{-0.6cm}
\end{figure*}

Our analysis is based on the hypothesis that there is a kind of universal curve relating $P_{RR}$ with the initial concurrence, as Fig. \ref{fig:5}(c) suggests. Let us start with the same initial state [Eq. (\ref{Initialdetection})] and express the qubit operators in the even and odd basis:
\begin{multline}\label{eodecomp}
\vert \Psi(0)\rangle = \left(\int dx \; \psi(x;x_0)c_R^\dagger(x)\right)\otimes \\ \otimes \left(\frac{(1+\xi)\sigma_e^\dagger + (1-\xi) \sigma_o^\dagger }{\sqrt{2}\sqrt{1+\vert\xi\vert^2}}\right)\vert 0\rangle.
\end{multline}
The contribution associated to the odd symmetry qubit state does not interact with the incident photon nor decays into the waveguide. Thus, as the total number of excitations ($N = 2$) is conserved, the time evolution of the term proportional to $\sigma_o^\dagger\vert 0 \rangle$ will correspond to a \textit{one-photon} state. On the other hand, the even term $\sigma_e^\dagger \vert 0 \rangle$ interacts with photons in a complex way, and will decay completely into photonic modes for sufficiently long times. As a consequence, at time $t \to \infty$ this term will evolve to a \textit{two-photon} state. Hence, by measuring any two-photon magnitude such as $P_{RR}$ we project onto the subspace proportional to $\sigma_e^\dagger \vert 0 \rangle$. In this way we eliminate the odd symmetry contribution, and the dependence with the parameter $\xi$ can be explicitly extracted out of $P_{RR}$:
\begin{equation}\label{PRR2}
\begin{split}
P_{RR} = \left\vert\frac{1+\xi}{\sqrt{2}\sqrt{1+\vert\xi\vert^2}} \right\vert^2\lim_{t\to\infty} \int_0^\infty dx_1\int_{x_1}^\infty dx_2 \\\bigg\vert c_R(x_1)c_R(x_2)\int dx c_R^\dagger(x) \psi(x;x_0)\vert 0\rangle \bigg\vert^2.
\end{split}
\end{equation}

Note that in the expression above, the dependence with all the system parameters except $\xi$ is contained inside the integrals. Finally, we introduce the two-photon detection probability for an initially disentangled state, $P_{RR}(\xi = 0)$, and rearrange the previous equality to obtain
\begin{equation}\label{final}
\left\vert\frac{P_{RR}(\xi)}{P_{RR}(\xi = 0)}-1\right\vert = \frac{2Re[\xi]}{1+\vert \xi \vert^2}.
\end{equation}
If the parameter $\xi$ is a real number, the expression above is equal to the concurrence [Eq. (\ref{ConcXI})], and thus the measurement of the normalized probability $P_{RR}(\xi)/P_{RR}(\xi = 0)$ fully determines the initial entanglement between the qubits. This situation is represented in Fig. \ref{fig:6}, where the relation between concurrence (dashed line) and normalized probability (solid line) is shown. On the other hand, when $\xi$ has an imaginary part, Eq. (\ref{final}) is less restrictive as we have $C(\xi) > \vert P_{RR}(\xi)/P_{RR}(\xi = 0) -1\vert$. This expression is still useful as it provides a lower bound for the concurrence of the qubits. The possibility of measuring entanglement only through two scattering measurements, $P_{RR}(\xi)$ and $P_{RR}(0)$, is a very powerful result. Note that, as Eq. (\ref{final}) holds for any two-photon probability, the same information can be obtained by measuring either $P_{LL}$ or $P_{RL}$. Moreover, this property is independent on all the system parameters, as it is valid for any qubits-waveguide coupling $\gamma$ and qubits transition frequency $\Omega$. Additionally, it does not depend on the single photon source, as the shape of the wavepacket is completely arbitrary (provided that it has a steep increasing wavefront). The inclusion of losses in the qubits due to emission into free-space modes would not modify this equality. The generality of our result allows for an easier experimental detection of entanglement in waveguide QED. 

\section{Conclusion}

Waveguide QED is a currently expanding field with numerous potential applications in quantum information and computation. The interaction between qubits coupled to a waveguide can be mediated by guided photons and it is possible to harness these processes to control the entanglement of the qubits. In this context we have explored three fundamental aspects in this work, namely entanglement generation, manipulation, and detection. Previous research has addressed entanglement control by pumping the qubits with an external source. Here we take advantage of the presence of the waveguide not only to couple the qubits to each other but also to excite them. The scattering of one or two guided photons impinging on the two qubit system gives rise to the entanglement of the pair, and we show that the corresponding time profile of the generated concurrence can be altered by adjusting the delay between both photons. Such modification is a consequence of sudden death and revival of entanglement phenomena taking place in the system. In a complementary way we also demonstrate that, given an initial one-excitation state of the qubits, it is possible to infer the initial concurrence of the two-qubit system by analyzing the scattering output of one photon impinging on them. Our results could have implications for devising scalable quantum devices, where waveguides have been suggested as ideal platforms for control and information transfer between different parts of a quantum network \cite{KimbleNATURE2008,LuPRA2014,ZhengPRL2013}. One of the key requirements for scalability purposes is the precise control of the qubit-qubit entanglement in waveguide systems, which could be achieved with the generation, manipulation, and detection schemes presented in this work. Moreover, the recent experimental advances in superconducting waveguides \cite{vanLooSCIENCE2013} show that waveguide QED is a realistic and promising candidate for quantum information purposes.

\begin{acknowledgements}
This work has been funded by the European Research Council (ERC-2011-AdG Proposal No. 290981) and by the
Spanish MINECO under contract MAT2011-28581-C02-01.
\end{acknowledgements}

\appendix*
\section{diagonalization of the hamiltonian in the $N=2$ Fock subspace}

This Appendix is devoted to  the diagonalization of the Hamiltonian in Eq.~(\ref{HEO}) in the two-excitation subspace. In order to do this, we start by rewriting this expression as
\begin{equation}\label{H=He+Ho}
H = H_e + H_{o1}+H_{o2},
\end{equation}
where the three components are defined as
\begin{equation}\label{He}
\begin{aligned}
H_e &= -iv_g\int dx c_e^\dagger(x)\partial_xc_e(x) + \\
+ &\Omega\sigma_e^\dagger\sigma_e  + V\int dx \delta(x) \left[c_e^\dagger(x)\sigma_e + \text{h.c.}\right],
\end{aligned}
\end{equation}
\begin{equation}
H_{o1}  =  -iv_g\int dx c_o^\dagger(x)\partial_xc_o(x),
\end{equation}
\begin{equation}
 H_{o2} =  \Omega\sigma_o^\dagger\sigma_o.
\end{equation}
The first Hamiltonian describes the two qubits coupled to the even parity photons. The second and third terms describe the free, odd parity photons and the uncoupled odd qubit state, respectively. It can be checked that the following commutation relations hold: 
\begin{equation}\label{commutation}
\left[H_e,H_{o1}\right] = \left[H_e,H_{o2}\right] = \left[H_{o1},H_{o2}\right] = 0.
\end{equation}
As a consequence, we can diagonalize the three different subspaces separately. The single-excitation eigenstates of $\lbrace H_{e},H_{o1},H_{o2}\rbrace$ have been discussed in detail in Ref.  \cite{GonzalezBallesteroNJPHYS2013}, and will be respectively labeled as $\lbrace \vert \epsilon_{e} \rangle_{s},\vert \epsilon_{o1} \rangle_{s},\vert \Omega_{o2} \rangle_{s}\rbrace$. It is straightforward to notice that $\vert \epsilon_{o1} \rangle_{s}$ is a free plane wave, while $\vert \Omega_{o2} \rangle_{s} = \vert - \rangle = \sigma_o^\dagger\vert 0\rangle$. The eigenstate $\vert \epsilon_{e} \rangle_{s}$ contains both a photonic part and a qubit contribution.

In the two-excitation subspace, we can construct eigenstates in two different ways: first, by taking the direct product of single-excitation states belonging to two different subspaces. This yields three eigenstates corresponding to the three possible direct products of $\lbrace \vert \epsilon_{e} \rangle_{s},\vert \epsilon_{o1} \rangle_{s},\vert \Omega_{o2} \rangle_{s}\rbrace$:
\begin{eqnarray}
\vert \epsilon_1\rangle =\vert k_{1,e}\rangle_s \otimes \vert k_{2,o1}\rangle_s ,
\\
\vert \epsilon_2\rangle =\vert (\epsilon-\Omega)_{e}\rangle_s \otimes \vert -\rangle ,
\\
\vert \epsilon_3\rangle =\vert (\epsilon-\Omega)_{o1}\rangle_s \otimes \vert -\rangle ,
\end{eqnarray}
where the energy is given by $\epsilon = v_g \left(k_1+k_2\right)$. 
The commutation relations (\ref{commutation}) guarantee that these direct products are eigenstates of the total Hamiltonian [Eq. (\ref{HEO})]. The second kind of eigenstates are formed by two excitations which belong to the same subspace. In the case of the two-level Hamiltonian $H_{o,2}$, the double occupation of the excited state is not possible, so no eigenstate of this kind exists. In the case of the free Hamiltonian $H_{o,1}$, the absence of interaction term makes the eigenstate to be also a direct product, $\vert k_{1,o1}\rangle_s \otimes \vert k_{2,o1}\rangle_s$. Additionally, the bosonic character of the photons has to be taken into account via symmetrization of the wavefunction, so the final expanded state reads
\begin{multline}
\vert \epsilon_4\rangle = \int dx_1 \int dx_2 \left(e^{ik_1x_1+ik_2x_2}+e^{ik_1x_2+ik_2x_1}\right) \\c_o^\dagger(x_1)c_o^\dagger(x_2)\vert 0 \rangle.
\end{multline}
The fifth and last state is the most laborious to obtain. It is formed by two excitations in the subspace spanned by $H_e$, and cannot be expressed as a direct product. We follow the procedure of Ref. \cite{BarangerPRA2010} and make the following Ansatz for this eigenstate:
\begin{multline}
\vert\epsilon_5\rangle = \int dx_1 \int dx_2 \phi_{ee}(x_1,x_2) c^\dagger_e(x_1)c^\dagger_e(x_2)\vert 0 \rangle + \\\int dx \alpha(x) c^\dagger_e(x)\sigma_e^\dagger\vert 0 \rangle+\beta \sigma_e^\dagger\sigma_e^\dagger\vert 0 \rangle.
\end{multline}
The next step is to apply the even Hamiltonian (\ref{He}) over this state and set the time-independent Schr\"odinger equation, $H_e\vert \epsilon_5\rangle = \epsilon\vert\epsilon_5\rangle $. We can then split the result into independent components, arriving to the following system of equations:
\begin{equation}\label{eqAnotas}
(\epsilon - 2\Omega)\beta = V\alpha(0),
\end{equation}

\begin{equation}
(\epsilon - \Omega +iv_g\partial_x) \alpha(x) = V\beta\delta(x) + 2V\phi_{ee}(0,x),
\end{equation}
\begin{multline}
(\epsilon +iv_g\partial_1+iv_g\partial_2) \phi_{ee}(x_1,x_2) = \\ = \frac{V}{2}\left[\delta(x_1)\alpha(x_2)+\delta(x_2)\alpha(x_1)\right].
\end{multline}
To get rid of the delta functions, we integrate the equations in the vicinities of $x=0$. After manipulating the result, we arrive to the following equivalent system:
\begin{equation} \label{aa}
(\epsilon - 2\Omega)\beta = V\alpha(0),
\end{equation}
\begin{equation}\label{bb}
iv_g(\phi_{ee}(0^+,x) -\phi_{ee}(0^-,x))= \frac{V}{2} \alpha (x),
\end{equation}
\begin{equation}\label{cc}
(\epsilon +iv_g\partial_1+iv_g\partial_2) \phi_{ee}(x_1,x_2) = 0,
\end{equation}
\begin{equation}\label{dd}
iv_g(\alpha(0^+) - \alpha(0^-)) = V\beta,
\end{equation}
\begin{multline}\label{ee}
\left(\epsilon-\Omega +i\frac{\gamma}{2}+iv_g\partial_x\right)\phi_{ee}(0^+,x) = \\= \left(\epsilon-\Omega -i\frac{\gamma}{2}+iv_g\partial_x\right)\phi_{ee}(0^-,x).
\end{multline}
The next step is the following: first, we use Eqs. (\ref{cc}) and (\ref{dd}) to solve for the two-photon wavefunction $\phi(x_1,x_2)=\phi(x_2,x_1)$ (the symmetry under the permutation $\lbrace x_1\leftrightarrow x_2\rbrace$ obeys the bosonic character of photons). This has been done in detail in Ref. \cite{BarangerPRA2010}, where the wavefunction is determined with the exception of a free parameter $C$. We then have a system of three equations (\ref{aa}, \ref{bb}, and \ref{dd}) for the three unknowns $\alpha(x),\beta,C$. The final solutions are more easily written as a function of the coefficients $c_j^\pm=k_j - \left(\Omega/v_g\right) \pm i\left(\gamma/2v_g\right)$, and are given by
\begin{equation}\label{beta}
\beta = \frac{2\gamma/v_g}{ c_1^+ c_2^+}\frac{c_1^+ + c_2^+}{c_1^+ + c_2^+-i\gamma/2v_g},
\end{equation}
\begin{widetext}
\begin{equation}\label{alpha}
\alpha(x) = 2\sqrt{\frac{\gamma}{v_g}} \cdot \left\{\begin{array}{ll}
\frac{e^{ik_1x}}{c_2^+} + \frac{e^{ik_2x}}{c_1^+} & \text{for} \;\;\;\;  x<0, \\ 
\\
\frac{1}{c_1^+} \frac{1}{c_2^+} \Big[c_2^-e^{ik_2x}+c_1^-e^{ik_1x} + i\frac{\gamma}{v_g}\frac{c_1^+ + c_2^-}{c_1^+ + c_2^+ -i\gamma/2v_g}e^{i(k_1+k_2)x}e^{(-i-(\gamma/2v_g))x}\Big] & \text{for} \;\;\;\;  x>0. 
\end{array}\right.
\end{equation}


\begin{equation}\label{Phi}
\phi(x_1,x_2) = \left\{\begin{array}{ll}
e^{ik_1x_1+ik_2x_2} + e^{ik_1x_2+ik_2x_1} & \text{for} \;\;\;\; x_1<x_2<0, \\
\\
\frac{c_2^{-}}{c_2^{+}}e^{ik_1x_1+ik_2x_2} + \frac{c_1^{-}}{c_1^{+}}e^{ik_1x_2+ik_2x_1} &  \text{for} \;\;\;\; x_1<0<x_2, \\
\\
\frac{c_1^{-}}{c_1^{+}}\frac{c_2^{-}}{c_2^{+}} \left(e^{ik_1x_1+ik_2x_2} + e^{ik_1x_2+ik_2x_1}\right) + \\+\theta(x_2-x_1) \frac{\left(\gamma/v_g\right)^2}{c_1^{+}c_2^{+}} \frac{c_1^{+}+c_2^{-}}{c_1^{+}+c_2^{+}-i\gamma/2v_g} e^{i \Omega x_1/v_g}e^{i(\epsilon-\Omega)x_2/v_g} e^{-\gamma\vert x_2-x_1\vert/2v_g} & \text{for} \;\;\;\; 0<x_1<x_2.
\end{array}\right.
\end{equation}

\end{widetext}
 Let us focus on interpreting the two-photon wavefunction (Eq. \ref{Phi}). In the region $x_1,x_2<0$ it is a boson-symmetrized plane wave; then, when one of the photons crosses the boundary $x=0$, each term is multiplied by the single-photon transmission coefficient \cite{GonzalezBallesteroNJPHYS2013}. Finally, when both photons have interacted with the qubits, two contributions appear in the wavefunction: the first one is the linear term, in which the incoming wave is just multiplied by the transmission coefficient for both photons. These linear terms are responsible for the two excitation transitions we discuss in Fig. \ref{fig:4} (a), via the usual reflection and transmission processes. The second term in the transmitted wavefunction is a two-photon bound state coming from the discrete nature of the qubits' energy levels. This process has been studied in detail in previous works \cite{FanPRL2007,ShiPRA2011,HoffmannPRA20033,RephaeliPRA2011}, and is responsible for the photon-induced decay of a previously excited qubit into the waveguide, i.e., the stimulated emission shown in Fig. \ref{fig:4} (a). Finally, the value of $\phi(x_1,x_2)$ in the region $x_1 > x_2$ (note that Eq. (\ref{Phi}) shows the solution only in $x_1<x_2$) can be obtained by using the boson symmetry \cite{BarangerPRA2010}.

After the eigenstates have been explicitly calculated, the time-evolution operator can be obtained. Taking into account the bosonic symmetry and the fact that $\vert \epsilon_4\rangle$ and $\vert \epsilon _5\rangle$ are not normalized to unity, we arrive to the following compact expression:
\begin{equation}
\begin{aligned}
U(t) & = \frac{1}{2\pi v_g} \int d\epsilon e^{-i\epsilon t} \left(\vert \epsilon_2\rangle \langle \epsilon_2\vert+\vert \epsilon_3\rangle \langle \epsilon_3\vert\right) +\\ + & \frac{1}{2}\frac{1}{16\pi^2}\int dk_1\int dk_2 e^{-iv_g\left(k_1+k_2\right) t} \big(\vert\epsilon_5\rangle\langle\epsilon_5\vert+ \\+& \vert\epsilon_4\rangle\langle\epsilon_4\vert+8\vert\epsilon_1\rangle\langle\epsilon_1\vert\big).
\end{aligned}
\end{equation}
where we have taken $\Omega = 1$ for simplicity.

\end{document}